\documentstyle[12pt]{article}

\expandafter\ifx\csname mathrm\endcsname\relax\def\mathrm#1{{\rm #1}}\fi

\makeatletter
\if@twoside \oddsidemargin -17pt \evensidemargin 00pt
\else \oddsidemargin 00pt \evensidemargin 00pt
\fi
\expandafter\ifx\csname mathrm\endcsname\relax\def\mathrm#1{{\rm #1}}\fi

\makeatother

\def\beq{\begin{equation}}
\def\eeq{\end{equation}}
\def\beqar{\begin{eqnarray}}
\def\eeqar{\end{eqnarray}}
\def\barr#1{\begin{array}{#1}}
\def\earr{\end{array}}
\def\bfi{\begin{figure}}
\def\efi{\end{figure}}
\def\btab{\begin{table}}
\def\etab{\end{table}}
\def\bce{\begin{center}}
\def\ece{\end{center}}
\def\nn{\nonumber}
\def\disp{\displaystyle}
\def\text{\textstyle}

\def\al{\alpha}

\def\ga{\gamma}
\def\de{\delta}
\def\eps{\varepsilon}
\def\la{\lambda}
\def\si{\sigma}
\def\ie{i\epsilon}
\def\Ga{\Gamma}
\def\De{\Delta}

\def\gWpm{g_{\PWpm}^2(0)}
\def\gWO{g_{\PWO}^2(\MZ^2)}

\def\refeq#1{\mbox{(\ref{#1})}}
\def\reffi#1{\mbox{Fig.~\ref{#1}}}
\def\reffis#1{\mbox{Figs.~\ref{#1}}}
\def\refta#1{\mbox{Tab.~\ref{#1}}}
\def\refse#1{\mbox{Sect.~\ref{#1}}}
\def\refses#1{\mbox{Sects.~\ref{#1}}}
\def\refapp#1{\mbox{App.~\ref{#1}}}
\def\citere#1{\mbox{Ref.~\cite{#1}}}
\def\citeres#1{\mbox{Refs.~\cite{#1}}}

\newcommand{\GeV}{\unskip\,\mathrm{GeV}}
\newcommand{\MeV}{\unskip\,\mathrm{MeV}}
\newcommand{\TeV}{\unskip\,\mathrm{TeV}}

\newcommand{\nba}{\unskip\,\mathrm{nb}}
\renewcommand{\O}{{\cal O}}

\def\mathswitchr#1{\relax\ifmmode{\mathrm{#1}}\else$\mathrm{#1}$\fi}

\newcommand{\PW}{\mathswitchr W}
\newcommand{\PZ}{\mathswitchr Z}

\newcommand{\PH}{\mathswitchr H}
\newcommand{\Pe}{\mathswitchr e}

\newcommand{\Pd}{\mathswitchr d}
\newcommand{\Pf}{\mathswitchr f}
\newcommand{\Ph}{\mathswitchr h}
\newcommand{\Pl}{\mathswitchr l}
\newcommand{\Pu}{\mathswitchr u}
\newcommand{\Ps}{\mathswitchr s}
\newcommand{\Pc}{\mathswitchr c}
\newcommand{\Pb}{\mathswitchr b}
\newcommand{\Pt}{\mathswitchr t}
\newcommand{\Pq}{\mathswitchr q}
\newcommand{\Pep}{\mathswitchr {e^+}}
\newcommand{\Pem}{\mathswitchr {e^-}}

\newcommand{\PWp}{\mathswitchr {W^+}}
\newcommand{\PWm}{\mathswitchr {W^-}}
\newcommand{\PWpm}{\mathswitchr {W^\pm}}
\newcommand{\PWO}{\mathswitchr {W^0}}
\newcommand{\PZO}{\mathswitchr {Z^0}}

\newcommand{\Zbb}{$\PZO\to\Pb\bar\Pb$}
\newcommand{\Zff}{$\PZO\to\Pf\bar\Pf$}
\newcommand{\Zqq}{$\PZO\to\Pq\bar\Pq$}
\newcommand{\Zuudd}{$\PZO\to\Pu\bar\Pu,\Pd\bar\Pd$}
\newcommand{\Znn}{$\PZO\to\nu\bar\nu$}

\newcommand{\Gb}{\Ga_\Pb}
\newcommand{\Gd}{\Ga_\Pd}
\newcommand{\Gu}{\Ga_\Pu}
\newcommand{\Gq}{\Ga_\Pq}
\newcommand{\Gh}{\Ga_{\rm h}}
\newcommand{\GT}{\Ga_{\rm T}}
\newcommand{\Gl}{\Ga_{\rm l}}

\def\mathswitch#1{\relax\ifmmode#1\else$#1$\fi}

\newcommand{\MW}{\mathswitch {M_\PW}}
\newcommand{\MWpm}{\mathswitch {M_\PWpm}}
\newcommand{\MWO}{\mathswitch {M_\PWO}}

\newcommand{\MZ}{\mathswitch {M_\PZ}}
\newcommand{\MH}{\mathswitch {M_\PH}}

\newcommand{\Mb}{\mathswitch {m_\Pb}}
\newcommand{\Mt}{\mathswitch {m_\Pt}}

\newcommand{\sw}{\mathswitch {s_\PW}}

\newcommand{\swbar}{\mathswitch {\bar s_\PW}}
\newcommand{\swfbar}{\mathswitch {\bar s_{\PW,\Pf}}}
\newcommand{\swqbar}{\mathswitch {\bar s_{\PW,\Pq}}}

\newcommand{\GF}{\mathswitch {G_\mu}}

\newcommand{\yb}{y_\Pb}
\newcommand{\yq}{y_\Pq}
\newcommand{\yu}{y_\Pu}
\newcommand{\yd}{y_\Pd}
\newcommand{\logx}{f_1}

\newcommand{\alpz}{\alpha(\MZ^2)}
\newcommand{\alps}{\alpha_{\rm s}}
\newcommand{\onel}{{\mathrm{1-loop}}}
\newcommand{\bos}{{\mathrm{bos}}}
\newcommand{\fer}{{\mathrm{ferm}}}
\newcommand{\uni}{{\mathrm{univ}}}
\newcommand{\BW} {{\mathrm{WPD}}}
\newcommand{\BZ} {{\mathrm{ZPD}}}

\newcommand{\yh}{y_{\rm h}}
\newcommand{\mass}{{\mathrm{mass}}}
\newcommand{\QCD}{{\mathrm{QCD}}}
\newcommand{\LEP}{{\mathrm{LEP}}}
\newcommand{\SLD}{{\mathrm{SLD}}}

\def\Li{\mathop{\mathrm{Li}_2}\nolimits}
\def\Re{\mathop{\mathrm{Re}}\nolimits}

\def\draftdate{\relax}
\def\mda{\relax}
\def\mua{\relax}
\def\mla{\relax}
\def\draft{
\def\thtystars{******************************}
\def\sixtystars{\thtystars\thtystars}
\typeout{}
\typeout{\sixtystars**}
\typeout{* Draft mode!
         For final version remove \protect\draft\space in source file *}
\typeout{\sixtystars**}
\typeout{}
\def\draftdate{\today}
\def\mua{\marginpar[\boldmath\hfil$\uparrow$]%
                   {\boldmath$\uparrow$\hfil}%
                    \typeout{marginpar: $\uparrow$}\ignorespaces}
\def\mda{\marginpar[\boldmath\hfil$\downarrow$]%
                   {\boldmath$\downarrow$\hfil}%
                    \typeout{marginpar: $\downarrow$}\ignorespaces}
\def\mla{\marginpar[\boldmath\hfil$\rightarrow$]%
                   {\boldmath$\leftarrow $\hfil}%
                    \typeout{marginpar: $\leftrightarrow$}\ignorespaces}
\overfullrule 5pt
\oddsidemargin -15mm
\marginparwidth 29mm
}

\makeatletter

\def\eqnarray{\stepcounter{equation}\let\@currentlabel=\theequation
\global\@eqnswtrue
\global\@eqcnt\z@\tabskip\@centering\let\\=\@eqncr
$$\halign to \displaywidth\bgroup\hskip\@centering
  $\displaystyle\tabskip\z@{##}$\@eqnsel&\global\@eqcnt\@ne
  \hskip 2\arraycolsep \hfil${##}$\hfil
  &\global\@eqcnt\tw@ \hskip 2\arraycolsep $\displaystyle\tabskip\z@{##}$\hfil
   \tabskip\@centering&\llap{##}\tabskip\z@\cr}
\def\appendix{\par
 \setcounter{section}{0} \setcounter{subsection}{0}
 \def\thesection{\Alph{section}}}

\makeatother

\begin{document}
\thispagestyle{empty}
\def\thefootnote{\fnsymbol{footnote}}
\setcounter{footnote}{1}
\null
\hfill BI-TP 94/62 \\
\null
\hfill hep-ph/9501404
\vskip .8cm
\begin{center}
{\Large \bf Refined Analysis of the Electroweak Precision Data%
\footnote{Partially supported by the EC-network contract CHRX-CT94-0579.}
\par}
\vskip 3em
{\large S.\ Dittmaier%
\footnote{Supported by the Bundesministerium f\"ur Forschung
und Technologie, Bonn, Germany.}%
, D.\ Schildknecht}
\vskip .5em
{\it Fakult\"at f\"ur Physik, Universit\"at Bielefeld, Germany}
\vskip 2em
{\large M.\ Kuroda}
\vskip .5em
{\it Department of Physics, Meiji-Gakuin University, Yokohama, Japan}
\end{center} \par
\vskip 1.2cm
\vfil
{\bf Abstract} \par
We refine our recent analysis of the electroweak precision data at the
\PZO\ pole by including the hadronic decay modes of the \PZO. Within the
framework of an effective Lagrangian we parametrize $SU(2)$ violation by
the additional process-specific parameters $\De y_\nu$, $\De\yh$,
and $\De\yb$ (for the $\PZO\nu\bar\nu$, $\PZO\Pq\bar\Pq$, and
$\PZO\Pb\bar\Pb$ vertices) together with the previously introduced parameters
$\De x$, $\De y$, and $\eps$. We find that a six-parameter analysis of
the experimental data is indeed feasible, and it is carried out in
addition to a four-parameter fit for $\De x$, $\De y$, $\eps$, and
$\De\yb$ only. We reemphasize that the experimental
data have become sensitive to the (combined) magnitude of the vertex
corrections at the $\PWp\Pl\bar\nu$ ($\PWm\nu\bar\Pl$) and $\PZO\Pl\bar\Pl$
vertices, $\De y$, which is insensitive to the notion of the Higgs
mechanism but dependent on the non-Abelian, trilinear vector-boson
coupling. Full
explicit analytical results for the standard one-loop predictions
for the above-mentioned parameters
are given, and the leading two-loop top-quark effects are included.
The analytic formluae for the analysis of the experimental data in terms
of the parameters $\De x$, $\De y$ etc.\ are
presented in order to encourage experimentalists to persue such an
analysis by themselves with future data.
\par
\vskip 1cm
\noindent December 1994 \par
\null
\setcounter{page}{0}
\clearpage
\def\thefootnote{\arabic{footnote}}
\setcounter{footnote}{0}

\section{Introduction}
\label{intro}

Based on an effective Lagrangian \cite{bi93} which parameterizes
$SU(2)$ violation in terms of three parameters, $\De x, \De y, \eps$,
we have recently presented an analysis \cite{di94} of the experimental
data \cite{mordat} on the leptonic width,
$\Gl$, of the \PZO, the leptonic mixing angle, $\swbar^2$,
and the \PWpm\ mass, $\MWpm$. In systematically
discriminating fermion-loop corrections to the \PWpm, $\ga$, \PZ\
propagators from all the other one-loop corrections which
depend on the empirically unknown non-Abelian couplings and the
properties of the Higgs scalar,
we obtained the striking result
that contributions beyond the pure fermion-loop corrections are required
for consistency with the experimental data.
This is a consequence of the high accuracy reached by the most recent data.
More specifically, we found that
such bosonic contributions are required in the parameter $\De y$, which
connects the \PWpm--fermion-coupling (determined via $\mu^\pm$ decay) with
the \PWO--fermion-coupling at the \PZO-mass shell,
$\gWpm \equiv 4\sqrt{2}G_\mu\MW^2 = (1+\De y) \gWO$.
Within the standard model, the bosonic contribution to $\De y$ is practically
independent of the Higgs mass, $\MH$, but contains vertex
corrections to the couplings of the \PWpm\ and the \PZO\ to fermions
which depend on the trilinear vector-boson self-couplings:
the experimental data have indeed become sensitive to bosonic loops,
i.e.\ to corrections
which involve the bosonic couplings of the vector bosons among each other.

In the present paper, we extend our previous results by including the
hadronic observables, the total \PZO\ width, $\GT$, as well as
the partial widths for \PZO\ decay into hadrons, $\Gh$, and for
\PZO\ decay into $\Pb\bar\Pb$ pairs, $\Gb$, in our analysis.
Accordingly, we generalize our effective Lagrangian to include vertex
modifications which are specific for the couplings of the \PZO\ to the
different quark flavors. This amounts to introducing the parameters
$\De y_{\Pu,\Pd} = \De y_{\Pc,\Ps}$ and $\De\yb$ (in addition to the
vertex modification at the \PZO\ $\nu\bar\nu$ vertex, $\De y_\nu$).
Explicit analytical expressions for these parameters in the standard
electroweak theory are given.

In conjunction with the inclusion of the hadronic observables, we also
update our previous analysis \cite{di94} by using the most recent
experimental data \cite{sc94}.

In the analysis of the data we will proceed in two steps. In a first step,
we present the results of a purely phenomenological six-parameter
analysis, determining the parameters $\De y_\nu$, a linear combination of
$\De\yu$ and $\De\yd$ to be called $\De\yh$, and $\De\yb$
(in addition to the ``leptonic'' parameters $\De x,\De y,
\eps$) from the above-mentioned six observables, $\MWpm$, $\Gl$, $\swbar^2$
and $\GT, \Gh, \Gb$. It is remarkable that such a six-parameter
analysis of the data is possible with reasonable experimental errors. In a
second step, we take advantage of the fact that $\De y_\nu$ and
$\De\yh$ can actually be reliably calculated in the standard model
solely from the
empirically well-known couplings of the vector-bosons to fermions.
The parameters $\De y_\nu$ and $\De\yh$ are indeed on the same footing
as the contributions $\De x^\fer$, $\De y^\fer$, $\eps^\fer$ to $\De x$,
$\De y$, $\eps$ resulting from
the fermion-loop corrections to the vector-boson propagators already
employed previously. The number of free parameters thus being reduced to
four, we will present fits to the mentioned six observables as well as
fits to five observables, by singling out and excluding $\Gb$.

A few general comments on our procedure seem appropriate:
\begin{itemize}
\item
It is sometimes argued that a direct comparison of the observables with
(standard) theoretical predictions may be sufficient, rather than
comparing with the parameters of an effective Lagrangian essentially
introduced by $SU(2)$-symmetry requirements. This is by no means true,
however. Indeed, while the necessity for bosonic corrections in addition
to fermion-loop propagator corrections became apparent in our direct
comparison with the observables $\MWpm$, $\swbar^2$, $\Gl$
(see Figs.~1--3 in \citere{di94}),
the detailed nature of these corrections only became clear {\it after the
transition to the parameters} $\De x,\De y,\eps$. More precisely,
it turned out that
approximating $\De x$ and $\eps$ by fermion loops only, $\De x \approx
\De x^\fer$, $\eps \approx \eps^\fer$, lead to agreement with
experiment. On the other hand, a non-zero contribution to $\De y$ in
addition to the fermion loops, $\De y^\fer$,
was shown to be necessary. In the standard theory it
is provided by the above-mentioned contribution to
$\De y$ which corresponds to (combined) vertex corrections at the \PZO\
and the \PWpm\ couplings to fermions and depends on the non-Abelian
structure of the theory.
However, it turned out that the experimental resolution of
$\De y$ does not yield any constraint with respect to the Higgs
mechanism. In fact, $\De y$ can be predicted in an electroweak massive
vector-boson theory \cite{di94a}.
\item
As in our previous work \cite{di94}, in clearly separating fermion-loop
effects to $\ga$ {\it as well as} \PZO\ and \PWpm\ propagation
(and additional effects in the present paper which are on the same
ground theoretically), we clearly differ from related work \cite{okun}
in which
the $\alpz$-Born approximation (i.e. fermion loops in {\it $\ga$-propagation
only}) is compared with the data and the full electroweak theory.
While such an analysis shows that contributions beyond the $\alpz$-Born
approximation are needed, it does not provide information about
the detailed nature of these additional contributions within electroweak
theory. With respect to the experimental data on the effective mixing
angle, $\swbar^2$, the effect of separating bosonic and fermionic loops
was also explored in \citere{ga94}.
The measured value of the \PWpm\ mass was focussed on in \citere{hi94} with
respect to the necessity of corrections beyond the $\alpz$-Born
approximation.
\item
We note that our work differs from \citeres{alta,ellis} in so far as our
emphasis is on precision tests of the bosonic sector of the electroweak
theory via fully separating fermionic and bosonic corrections, while the
main emphasis in \citeres{alta,ellis} is put on
testing extensions of the standard theory, such as supersymmetry
or technicolor.
\end{itemize}

In \refse{efflag}, we present the extended version of our effective
Lagrangian for the description of \PZO\ interactions at one-loop
level. In \refse{anresm}, we turn to the analytic results for the
parameters introduced in addition to the ones used previously,
referring to the appendix for a few technical details. In
\refse{otherrcs},
we briefly summarize (known) two-loop and QCD corrections relevant
for the analysis of the observables in \refse{anexda}. Final
conclusions are drawn in \refse{concl}.

\section{The effective Lagrangian}
\label{efflag}

\subsection{The leptonic sector}
\label{lepton}

We first of all consider the case of charged leptons%
\footnote{Lepton universality is assumed throughout in our effective
Lagrangian.}
interacting with
the charged and neutral vector bosons \PWpm, $\ga$ and \PZO.
We start by slightly refining the treatment \cite{bi93,di94}
of neutral-current
interactions in the leptonic sector in terms of an effective Lagrangian.
Starting from the charged-current interaction in
\beq
{\cal L}_{\rm C}=-{1\over 2} W^{+\mu \nu} W^-_{\mu \nu} -
\frac{g_\PWpm}{\sqrt 2}\left( j^+_\mu W^{+\mu} + h.c.\right)
+ \MWpm^2 W^+_\mu W^{-\mu},
\label{lc}
\eeq
$SU(2)$ symmetry is broken in the transition to neutral-current
interactions by introducing the parameters $x$ and $y$ via
\beq
\MWpm^2 = x\MWO^2 = (1+\De x) \MWO^2,
\label{mwpm}
\eeq
and
\beq
4\sqrt 2 G_\mu \MWpm^2 \equiv \gWpm = y \gWO = (1+\De y) \gWO,
\label{y}
\eeq
as well as mixing of strength $\la$ between the electromagnetic and
neutral \PWO\ fields,
\beq
{\cal L}_{\rm mix} = -\frac{\la}{2} A_{\mu \nu} W^{0,\mu \nu}.
\label{lm}
\eeq
As in \citere{bi93,di94}, the parameter $\la$ will be replaced by
$\varepsilon$ via%
\footnote{Our notation is an outgrowth of several steps which lead to
the underlying Lagrangian. The mixing parameter $\eps$ was first
introduced in \citere{kn91}, and $\De x$, $\De y$ in \citere{bi93}.}
\beq
\la\equiv {e(\MZ^2)\over{g_\PWO(\MZ^2)}} (1-\eps),
\label{la}
\eeq
where $g_\PWO$ refers to the coupling of a charged
lepton to the neutral member of the (\PWpm,\PWO)-triplet.
For definiteness, we have indicated the energy scales as arguments of the
various couplings.
The electromagnetic coupling at the $\PZO$ mass has been denoted by
$e(\MZ^2)$ with
\beq
\frac{e^2(\MZ^2)}{4\pi} = \alpz \approx 1/129,
\eeq
where $\alpz$ includes the ``running'' of electromagnetic vacuum
polarization induced by the light fermions \cite{alpz1}.

A few additional remarks on the parameters introduced in (\ref{mwpm}) to
(\ref{la}) are appropriate and useful in connection with the realization of
non-zero values of
these parameters by loop corrections to be discussed in \refse{anresm}.
As $x$ in (\ref{mwpm}) is defined as a mass ratio, it is obviously a
``universal'' quantity, i.e., it is independent of the external
particles participating in a specific process
described by Lagrangian (\ref{lc}) and its neutral-current counterpart
to be defined below.
Likewise, the parameter $\la$ in (\ref{la}), as a mixing strength
among neutral boson fields, is a universal quantity. In contrast,
as $y$ in (\ref{y}) relates the charged current coupling deduced from the
specific process of $\mu^\pm$ decay to the coupling of the neutral
component of the (\PWpm,\PWO)-triplet to charged leptons, the
parameter $y$ obviously contains a process-specific contribution.
Assuming that $y$ is induced by loop corrections, as in the standard
theory to be discussed in \refse{anresm}, $y$ will also contain a
process-independent, i.e.\ universal part, as according to (\ref{y})
it relates quantities at different energy scales. Accordingly, in
linear expansion (\ref{y}) may be written as
\beq
g_\PWpm(0) = g_\PWO(\MZ^2) \left(1+\frac{1}{2} \De y^\BZ +
\frac{1}{2} \De y^\BW + \frac{1}{2} \De y^\uni\right),
\label{ysplit}
\eeq
where $\De y^\BZ$ and $\De y^\BW$ refer to the \PZO\ and \PWpm\
vertices, respectively, and $\De y^\uni$ denotes an additional
universal part.
As the expression (\ref{la}) for the universal (neutral-current)
mixing strength, $\la$, in terms of $\eps$ contains the process-specific
coupling $\gWO$, also $\eps$ in (\ref{la}) has to contain
a process-specific contribution,
\beq
\la={{e(\MZ^2)}\over{g_\PWO(\MZ^2)}}
\left( 1 - \eps^{\BZ} - \eps^{\uni}\right).
\label{la2}
\eeq
which must be identical to the \PZO-process-specific part in $\gWO$ in
(\ref{ysplit}), i.e.,
\beq
\eps^{\BZ} = {1\over 2} \De y^{\BZ}.
\label{epszpd}
\eeq
This relation will be seen to emerge also in \refse{anresm}, where the
expressions
for $\De x,\De y,\eps$ induced by loop effects in the $SU(2)\times U(1)$
theory will be given.

In order to construct the effective \PZO\ interaction, we still have
to specify the interaction of the photon and the interaction of the
third component of the (\PWpm,\PWO)-triplet with fermions. Noting that
the effective Lagrangian will have to contain standard one-loop
interactions as a special case, we couple the photon (at the \PZO-mass
scale) to a parity violating current via the replacement
\beq
A_\mu j^\mu_{\rm em} \to A_\mu Q
\left( \bar\psi_{\rm L} \ga^\mu \psi_{\rm L}
+ (1+\de)\bar\psi_{\rm R}\ga^\mu\psi_{\rm R}\right),
\label{delta}
\eeq
where obviously $\de$ has to vanish in the limit of real photons
interacting with fermions.

Skipping an explicit presentation of the \PWO\ interaction
\cite{bi93,di94}, upon
diagonalization, the \PZO\ part of the neutral-current Lagrangian
becomes,
\beqar
{\cal L}_{\rm N} & = &
-{1\over 4} Z_{\mu \nu} Z^{\mu \nu} +
\frac{\MWO^2}{2(1-\lambda^2)} Z_\mu Z^\mu
\nn\\
&& -\frac{g_\PWO}{\sqrt{1-\lambda^2}}
\left[j^{3}_\mu
-\frac{e\lambda}{g_\PWO}Q_\Pl
\left(\bar\psi_{\rm L} \ga_\mu \psi_{\rm L}
+(1+\de)\bar\psi_{\rm R} \ga_\mu \psi_{\rm R}\right)
\right] Z^\mu,
\label{lnc}
\eeqar
where $Q_\Pl=-1$ is the lepton charge. Introducing the following
linear combinations as auxiliary quantities,
\beq
x^{\prime} = x + 2 s^2_0 \de, \qquad
y^{\prime} = y - 2 s^2_0 \de, \qquad
\eps^{\prime}  = \eps - \de,
\label{xyeprime}
\eeq
and keeping $\de$ only in linear order ($\de$ in, e.g., the standard
electroweak theory turns out to be
extremely small, $\de\sim 10^{-4}$), the neutral-current Lagrangian
can be written as
\beqar
{\cal L}_{\rm N} & = &
-{1\over 4} Z_{\mu \nu} Z^{\mu \nu} +
\frac{\MWpm^2}{2x^{\prime}
\left(1-\swbar^2(1-\eps^{\prime})\right)}
Z_\mu Z^\mu   \nn\\[.3em]
&& -\frac{g_\PWpm}{\sqrt{y^{\prime}
\left(1-\swbar^2 (1-\eps^{\prime})\right)}}
\left[j^{3}_\mu - \swbar^2 j_{{\rm em},\mu}\right] Z^\mu.
\label{lnc2}
\eeqar
The (leptonic) weak mixing angle, $\swbar^2$, is empirically
determined by the charged lepton asymmetry at the \PZO\ resonance,
\beq
\swbar^2 = \frac{e^2(\MZ^2)}{g_\PWpm^2(0)}y^{\prime}(1-\eps^{\prime}),
\label{swbar}
\eeq
and $g_\PWpm\equiv g_\PWpm(0)$ is given in (\ref{y}).
$s^2_0$ in (\ref{xyeprime}) is defined via
\beq
s^2_0(1-s^2_0) = s_0^2c_0^2 = {{\pi \alpz}\over{\sqrt 2 G_\mu \MZ^2}}.
\label{s02}
\eeq

We note that Lagrangian (\ref{lnc2}) has the same form as
Lagrangian (10) in \citere{di94},
apart from the replacement of $x,y,\eps$ by $x^{\prime},y^{\prime},
\eps^{\prime}$. This replacement takes into
account a non-vanishing value of $\de$, which is in fact present in
the standard theory even though it is very small.
Nevertheless, from a principal point of view, it
has to be introduced in order to assure universality of $x$ in
(\ref{mwpm}).
In the subsequent analysis, standard model values for $\de$ will be
inserted.
This can be done without loss of generality for two reasons. First of
all, departures in $\de$ from its standard value according to
(\ref{xyeprime}) can always be absorbed
in the parameters $x,y,\eps$ to be determined from the experimental
data. Secondly, it will be seen that in the $SU(2)\times U(1)$ theory
$\de$ is on equal footing with the fermion-loop corrections and can be
calculated equally reliably.

Finally, we express the weak mixing angle, $\swbar^2$, the \PWpm\ mass,
$\MWpm$, and the leptonic \PZO\ width, $\Gl$, in terms of
$x^{\prime},y^{\prime},\eps^{\prime}$,
\beqar
\swbar^2 \left( 1-\swbar^2 \right)
& = &
\frac{\pi\alpz} {\sqrt 2 G_\mu \MZ^2}
\frac {y^{\prime}}{x^{\prime}} (1-\varepsilon^{\prime})
\frac{1} {\left( 1 +\frac{\bar\sw^2} {1-\bar\sw^2}
\varepsilon^{\prime} \right)}, \nn\\
\frac {\MWpm^2} {\MZ^2}
& = &
\left( 1-\bar\sw^2 \right)x^{\prime}
\left( 1 + \frac{\bar\sw^2}{1-\bar\sw^2}\varepsilon^{\prime} \right), \nn\\
\Gl
& = &
\frac {G_\mu \MZ^3} {24\pi \sqrt 2}
\left[ 1 + \left( 1 - 4\bar\sw^2\right)^2\right]
\frac{x^{\prime}}{y^{\prime}}
\left( 1 + \frac{3\alpha}{4\pi}\right).
\label{obslep}
\eeqar
Linearizing these observables also in $\De x,\De y,\eps$ yields
\beqar
\bar\sw^2 &=& s_0^2\left[1-\frac{1}{c_0^2-s_0^2}\varepsilon
-\frac{c_0^2}{c_0^2-s_0^2}(\Delta x-\Delta y)
+(c_0^2-s_0^2)\de
\right], \nn\\
\frac{\MWpm}{\MZ} &=& c_0\left[1+\frac{s_0^2}{c_0^2-s_0^2}\varepsilon
+\frac{c_0^2}{2(c_0^2-s_0^2)}\Delta x
-\frac{s_0^2}{2(c_0^2-s_0^2)}\Delta y
\right], \nn\\
\Gl &=& \Gl^{(0)}\left[1+
\frac{8s_0^2}{1+(1-4s_0^2)^2}\left\{
\frac{1-4s_0^2}{c_0^2-s_0^2}\varepsilon
+\frac{c_0^2-s_0^2-4s_0^4}{4s_0^2(c_0^2-s_0^2)}
(\Delta x-\Delta y)
+2s_0^2\de
\right\}\right], \hspace*{1.9em}
\label{obsleplin}
\eeqar
with
\beq
\Gl^{(0)} = \frac{\alpz\MZ}{48s_0^2c_0^2}
\left[1+(1-4s_0^2)^2\right]\left(1+\frac{3\alpha}{4\pi}\right).
\label{obsleplow}
\eeq
Obviously, upon absorbing $\de$ in $x,y,\eps$ according to
(\ref{xyeprime}), the relations (\ref{obsleplin}) agree with relations
(14) of \citere{di94} apart from the replacement of $x,y,\eps$ by
$x^{\prime},y^{\prime},\eps^{\prime}$.

\subsection{Generalization to arbitrary fermions}
\label{arbfer}

We turn to the generalization of the effective Lagrangian to the case
of neutrinos and, in particular, quarks. Accordingly, for each
flavor degree of freedom, we have to allow for a separate, process-specific
vertex correction. We define
\beq
\gWpm \equiv y y_\Pf g_{\PWO,\Pf}^2(\MZ^2) =
(1+\De y) (1+\De y_\Pf) g_{\PWO,\Pf}^2(\MZ^2),
\label{yf}
\eeq
where \Pf\ stands for any one of $\Pf=\nu,\Pu,\Pd,\Pc,\Ps,\Pb$.
Clearly, the case of $\De y_\Pf\equiv 0$ corresponds to the charged-lepton
case,
\beq
g_{W^0} \equiv g_{W^0, e, \mu, \tau}.
\label{gemt}
\eeq
In addition, we allow for the process-specific correction in the
electromagnetic current%
\footnote{Strictly speaking, we would also have to allow for a coupling
between the photon on the \PZO-mass scale and the neutrino. However, in
the \PZO\ part of the neutral-current Lagrangian this effect would merely
lead to a trivial redefinition of the parameter $\De y_\nu$.}
analogously to (\ref{delta}),
\beq
A_\mu j^\mu_{\rm em} \to A_\mu Q_\Pf
\left( \bar\psi_{\rm L} \ga^\mu \psi_{\rm L}
+ (1+\de+\de_\Pf)\bar\psi_{\rm R}\ga^\mu\psi_{\rm R}\right).
\label{deltaf}
\eeq
Passing through the diagonalization procedure previously employed,
we now have in generalization of (\ref{lnc})
\beqar
{\cal L}_{\rm N} & = &
-{1\over 4} Z_{\mu \nu} Z^{\mu \nu} +
\frac{\MWO^2}{2(1-\lambda^2)} Z_\mu Z^\mu
\nn\\
&& -\frac{y_\Pf g_\PWO}{\sqrt{1-\lambda^2}}
\left[j^{3}_\mu
-\frac{e\lambda}{y_\Pf g_\PWO}Q_\Pf
\left(\bar\psi_{\rm L} \ga_\mu \psi_{\rm L}
+(1+\de+\de_\Pf)\bar\psi_{\rm R} \ga_\mu \psi_{\rm R}\right)
\right] Z^\mu.
\label{lncf}
\eeqar
According to (\ref{lncf}), for each fermion flavor, $\Pf$, we have a specific
coupling strength and a specific mixing angle. For $y_\Pf = 1$,
$\de_\Pf=0$, we recover the lepton case (\ref{lnc}). In terms of a more
``physical'' set of parameters the Lagrangian (\ref{lncf}) takes the form
\beqar
{\cal L}_{\rm N} & = &
-{1\over 4} Z_{\mu \nu} Z^{\mu \nu} +
\frac{\MWpm^2}{2x^{\prime}
\left(1-\swbar^2(1-\eps^{\prime})\right)}
Z_\mu Z^\mu   \nn\\[.3em]
&& -\frac{g_\PWpm
\Big(1+2\vert Q_\Pf\vert s_0^2\de_\Pf
+2(\vert Q_\Pf\vert-1)s_0^2\de\Big)}
{\sqrt{y^{\prime}y_\Pf
\left(1-\swbar^2 (1-\eps^{\prime})\right)}}
\left[j^{3}_\mu - \swfbar^2 j_{{\rm em},\mu}\right] Z^\mu,
\label{lncf2}
\eeqar
where the weak mixing angle (for charged fermions) is given by
\beq
\swfbar^2 = \sqrt{y_\Pf}\,\swbar^2
+ s_0^2\left(1-2\vert Q_\Pf\vert s_0^2\right)\de_\Pf
+2(1-\vert Q_\Pf\vert)s_0^4\de.
\label{swf}
\eeq
The Lagrangian (\ref{lncf2}) generalizes (\ref{lnc2}) to the case of
an arbitrary fermion \Pf.
The fact that $\de_\Pf$ in (\ref{lncf2}) cannot be absorbed by a
redefinition of $y_\Pf$ in (\ref{lncf2}), (\ref{swf}) is of no practical
relevance.
Recall that the quark asymmetries depend on both the leptonic mixing
angle, $\swbar^2$, and the quark mixing angle, $\swqbar^2$ ($\Pf=\Pq$).
The dependence on $\swqbar^2$, however, is extremely weak so that it is
justified to replace $\swqbar^2$ by $\swbar^2$ in the quark asymmetries.
Consequently, $\de_\Pq$ cannot be separated from $\De\yq$ by measuring
the quark asymmetries. On the other hand,
also $\de_\Pf$ can be reliably calculated in the standard model.

With the modified Lagrangian (\ref{lncf2}) we obtain for the \Znn\ decay
width, $\Gamma_\nu$,
\beq
\Gamma_\nu = \frac {G_\mu \MZ^3} {12\pi \sqrt 2} \frac{x}{y\,y_\nu},
\label{obsnu}
\eeq
and for the \Zqq\ decay widths, $\Gq$,
\beqar
\Gq & = &
\frac {G_\mu \MZ^3} {8\pi \sqrt 2} \frac{x}{y\,y_\Pq}
\left[ 1 + \left( 1 - 4\vert Q_\Pq\vert\swqbar^2\right)^2\right]
\left( 1 + 4\vert Q_\Pq\vert s_0^2(\de+\de_\Pq)\right)
\left( 1 + Q_\Pq^2\frac{3\alpha}{4\pi}\right)
R_\QCD,
\hspace{2em}
\label{obshad}
\eeqar
where $\de$ and $\de_\Pq$ again are kept in linear order. Note that in
analogy to the $\O(\al)$ QED factor we have also applied the
QCD factor
\beq
R_{\QCD} = 1 + \left(\frac{\alps}{\pi}\right) +
1.41\,\left(\frac{\alps}{\pi}\right)^2 \,-\,
12.8\,\left(\frac{\alps}{\pi}\right)^3,
\eeq
to the hadronic widths.
$R_{\QCD}$ represents the one-, two- and three-loop QCD corrections
\cite{go91} corresponding to massless quarks. The inclusion of
finite-mass effects is described in \refse{otherrcs}.
For completeness, we write down the linearized form of the
observables in (\ref{swf}), (\ref{obsnu}) and (\ref{obshad}) yielding
\beq
\Gamma_\nu = \Gamma_\nu^{(0)}\left[1+\De x-\De y-\De y_\nu\right],
\label{obsnulin}
\eeq
for $\Gamma_\nu$ and
\beqar
\swqbar^2 &=& s_0^2\left[1-\frac{1}{c_0^2-s_0^2}\varepsilon
-\frac{c_0^2}{c_0^2-s_0^2}(\Delta x-\Delta y)
+\frac{1}{2}\De y_\Pq+(1-2s_0^2\vert Q_\Pq\vert)(\de+\de_\Pq)
\right], \nn\\[.3em]
\Gamma_\Pq &=& \Gamma_\Pq^{(0)}\left[1+
\frac{8s_0^2}{1+(1-4\vert Q_\Pq\vert s_0^2)^2}\left\{
\frac{\vert Q_\Pq\vert(1-4\vert Q_\Pq\vert s_0^2)}{c_0^2-s_0^2}\varepsilon
+\frac{2\vert Q_\Pq\vert s_0^2-1}{4s_0^2}\De y_\Pq
\right.\right.\nn\\[.2em]
&& \hspace{6em} \left.\left.
+\frac{c_0^2-s_0^2+4\vert Q_\Pq\vert s_0^4(1-2\vert Q_\Pq\vert)}
{4s_0^2(c_0^2-s_0^2)}(\Delta x-\Delta y)
+2s_0^2Q_\Pq^2(\de+\de_\Pq)
\right\}\right], \hspace*{1.9em}
\label{obshadlin}
\eeqar
for the hadronic observables. Here, we introduced the lowest-order decay
widths
\beqar
\Gamma_\nu^{(0)} &=& \frac{\alpz\MZ}{24s_0^2c_0^2}, \nn\\[.3em]
\Gamma_\Pq^{(0)}   &=& \frac{\alpz\MZ}{16s_0^2c_0^2}
\left[1+(1-4\vert Q_\Pq\vert s_0^2)^2\right]
\left( 1 + Q_\Pq^2\frac{3\alpha}{4\pi}\right)
R_\QCD.
\label{obsnuhadlow}
\eeqar

\section{Analytic results in the standard model}
\label{anresm}

In \citere{di94}, it was our essential point to systematically
distinguish between fermion-loop contributions {\it to $\ga$ as well
as \PWpm\ and \PZO\ propagation} (compare \reffi{ferloop}) and other
contributions, which in general depend on the couplings of the vector
bosons with each other.
\bfi
\begin{center}
\begin{picture}(16,3)
\put(-2.5,-14.0){\includegraphics{VV.ps}}
\end{picture}
\end{center}
\caption{Fermion-loop contributions to gauge-boson propagators.
\label{ferloop} }
\efi

As pointed out in the previous section, we first of all refine the
treatment of \citeres{bi93,di94} by also separating the
(numerically unimportant)
contribution called $\de$ in the previous section. This contribution
leads to a discrimination between right-handed and left-handed
fermions in the coupling to the photon
(compare (\ref{delta})) which in terms of \PZO\ couplings
corresponds to the diagrams shown in \reffi{delgra}. These diagrams form
a gauge-invariant set and only depend on the couplings of the bosons
to fermions. Accordingly, they are basically of similar
nature as the fermion loops in \reffi{ferloop} and can be reliably
calculated. The same conclusion will hold if the leptons
in \reffi{delta} are
replaced by quarks, thus yielding $\de_\Pq$. Both,
$\de$ and $\de_\Pq$ will be explicitly given below.
\bfi
\begin{center}
\begin{picture}(16,3.5)
\put(-2.5,-14.1){\includegraphics{delgra.ps}}
\end{picture}
\end{center}
\caption{Feynman graphs relevant for the parameter $\delta$.
\label{delgra} }
\efi

The second extension of our previous work is concerned with the vertex
correction $\De y$ for the case of quarks, $\De y_\Pq$, and the neutrino,
$\De y_\nu$. Since $\De y_\Pq$, $\De y_\nu$ are defined as deviations from
the leptonic parameter, $\De y$, all universal and WPD corrections are
 already included in $\De y$ so that $\De y_\Pq$, $\De y_\nu$ are entirely
furnished by ZPD contributions.
It is important to distinguish between the cases of the
light quarks $\Pq = \Pu, \Pd, \Pc, \Ps$ on the one hand, and the case of the
\Pb-quark on the other hand. The diagrams relevant for the
process-dependent contribution, $\De\yb$, are depicted in \reffi{dybgra}.
\begin{figure}
\begin{center}
\begin{picture}(16,11)
\put(-2.5,-6.3){\includegraphics{dybgra.ps}}
\end{picture}
\end{center}
\caption{Feynman diagrams relevant for the parameter $\De\yb$.}
\label{dybgra}
\efi
Note that the diagrams involving the Yukawa coupling of the charged
Goldstone scalar field, $\varphi$, to the fermion doublet (graphs
\reffi{dybgra}c),f),i),k),l),m) for the (\Pb,\Pt)-doublet) do not
contribute for
light doublets. Moreover, one finds that the contribution of the
diagram involving the non-Abelian vertex (i.e.\ the analogous ones
to graph \reffi{dybgra}j) for \Zbb) is already contained in $\De y$
for light doublets since these diagrams coincide with the one in the
lepton case (up to a necessary change in the sign if the charge flow
in the loop is inverted). Consequently, $\De y_\nu$ and $\De y_\Pq$ for
$\Pq=\Pu,\Pd,\Pc,\Ps$ do not contain Yukawa and
trilinear boson couplings, and
accordingly are again {\it on the same footing as the fermion loops}
in the gauge-boson propagators and can
be reliably calculated. The situation is different for \Zbb.
In this case, indeed, we obtain a contribution due
to Yukawa couplings and the trilinear non-Abelian vertex in conjunction
with the dependence on the mass of the top-quark, $\Mt$.

We turn to a representation of the analytic results. Separating fermion
loops (in $\ga, \PWpm, \PZO$ propagation) and the remaining contributions
which depend on the gauge-boson and the Higgs sector of the theory, as
in \citere{di94}, we have
\beq
a=a_{\fer} + a_{\bos},
\qquad {\rm with} \quad a=\De x, \De y, \eps,
\label{abosfer}
\eeq
and
\beq
a_{\bos} = a^{\uni}_{\bos} + a^{\BW} + a^{\BZ}.
\label{abospd}
\eeq
There is no reason to repeat the formulae given in \citere{di94}
apart from the (minor, theoretically relevant, but numerically
unimportant) change in the ZPD contributions due to the parameter
$\de$ introduced above.

It seems appropriate, however, to add a brief comment on the
general structure of our results before explicitly quoting them.
The (generally non-unique) splitting of the sum on the right-hand
side in (\ref{abospd}) into separately gauge-parameter-independent
vacuum-polarization
``univ'' and vertex (``WPD'', ``ZPD'') parts in \citere{di94} was
carried out by employing a procedure known as pinch technique
(see e.g.\ \citere{pt} and references therein).
It yields a somewhat different splitting in the right-hand side in
(\ref{abospd}) than other methods \cite{ke89,ku91}, and the results obtained
for the vertex functions coincide with the ones of the ``background
field method'' in the 't Hooft-Feynman gauge \cite{de94}.

The parameter $\De x$, defined by (\ref{mwpm}) is process-independent,
\beq
\De x^\uni_\bos\neq 0,\qquad \De x^\BZ = \De x^\BW = 0.
\label{deub}
\eeq
The only independent process-dependent parameters are contained in
the vertex modification $\De y$,
\beq
\De y^\uni_\bos\neq 0,\qquad \De y^\BZ\neq 0,\qquad \De y^\BW\neq 0,
\label{deub1}
\eeq
since the process-dependent part of $\eps$, is related to $\De y^\BZ$
via (compare \refeq{epszpd})
\beq
\eps^\uni_\bos\neq 0,\qquad \eps^\BZ =\frac{1}{2}\De y^\BZ,
\qquad \eps^\BW = 0.
\label{deub2}
\eeq

The explicit expressions for the universal and WPD contributions to
$\De x, \De y, \eps$ are exactly the ones given in \citere{di94}.
The ZPD parts are different owing to the introduction of $\de$,
\beq
\de=-\frac{\alpz s^2_0}{8\pi c^2_0}(11 + 16 C_1)
   =-\frac{\alpz s^2_0}{8\pi c^2_0}\left(11 - \frac{4}{3}\pi^2 \right)
   = 0.20\times 10^{-3}.
\label{de}
\eeq
The constants $C_{1,2,3}$ and the function $\logx(x)$ are defined in
\refapp{aux}.
The expression for $\eps^\BZ$ in the present notation is given by
\beqar
\varepsilon^\BZ &=&
\frac{\alpz}{4\pi s_0^2}\left[
(1-2s_0^2)^3\frac{2C_1}{c_0^2}
-(2-s_0^2)^2C_2-2c_0^4(3-s_0^2)C_3
\right.\nn\\ &&\phantom{\frac{\alpz}{24\pi s_0^2}}\left.
-\left(\frac{5}{2}-s_0^2\right)
\left(\log(c_0^2)-2c_0^2\logx(c_0^2)\right)
+\frac{17}{8c_0^2}-\frac{27s_0^2}{2c_0^2}+\frac{23s_0^4}{c_0^2}
-\frac{13s_0^6}{c_0^2}
\right], \hspace{2em}
\label{dzpd}
\eeqar
while $\De x^\BZ$ vanishes according to (\ref{deub}), and $\De y^\BZ$
follows from (\ref{deub2}) upon inserting $\eps^\BZ$ from (\ref{dzpd}).
We also note the numerical values of the
process-specific parameters
\beqar
\De y^\BZ &=& 2\eps^\BZ = 8.52 \times 10^{-3}, \nn\\
\De y^\BW &=& 5.46 \times 10^{-3},
\label{pdnum}
\eeqar
again referring to \citere{di94}
for the numerical values of the universal fermionic and bosonic
parts in the parameters $\De x, \De y, \eps$.

We turn to the standard values for the process-specific corrections,
$\de_\Pf$ and $\De y_\Pf$ ($\Pf=\nu,\Pu,\Pc,\Pd,\Ps,\Pb$).
The parameters $\de_\Pf$ are simply related to $\de$ and given by
\beq
\de_\Pf = (Q_\Pf^2-1)\de = (Q_\Pf^2-1)\times 0.20\times 10^{-3}.
\eeq
For fermions \Pf\ with light
isospin partners, $\De y_\Pf$ is obviously constant,
\beqar
\De y_\nu & = & \frac{\alpz}{8\pi c_0^2}\left[
16(3-6s_0^2+4s_0^4)C_1
+8c_0^2(2-s_0^2)^2C_2
+4c_0^2(5-2s_0^2)\log(c_0^2)
\right.\nn\\
&&\hphantom{\frac{\alpz}{8\pi c_0^2}}\left.
+55-96s_0^2+52s_0^4
\right], \nn\\[.3em]
\De\yu = \De y_\Pc & = & \frac{\alpz}{24\pi c_0^2}\left[
\frac{16}{9}(27-90s_0^2+76s_0^4)C_1
+8c_0^2(2-s_0^2)^2C_2
+4c_0^2(5-2s_0^2)\log(c_0^2)
\right.\nn\\
&&\hphantom{\frac{\alpz}{8\pi c_0^2}}\left.
+55-140s_0^2+\frac{908}{9}s_0^4
\right], \nn\\[.3em]
\De\yd = \De y_\Ps & = & \frac{\alpz}{12\pi c_0^2}\left[
\frac{16}{9}(27-72s_0^2+52s_0^4)C_1
+8c_0^2(2-s_0^2)^2C_2
+4c_0^2(5-2s_0^2)\log(c_0^2)
\right.\nn\\
&&\hphantom{\frac{\alpz}{8\pi c_0^2}}\left.
+55-118s_0^2+\frac{664}{9}s_0^4
\right],
\label{dyud}
\eeqar
or numerically,
\beqar
\De y_\nu & = & -3.05 \times 10^{-3}, \nn\\
\De\yu    & = & -0.82 \times 10^{-3}, \nn\\
\De\yd    & = & -1.82 \times 10^{-3}.
\eeqar
However, $\De\yb$ gets the above-mentioned
top-mass dependent contributions via virtual
W exchange. Proceeding analogously to our presentation \cite{di94} of
$\De x, \De y, \eps$, we split $\De\yb$ into a ``dominant'' $(dom)$
and a ``remainder'' $(rem)$ term,
\beq
\De\yb = \De\yb(dom) + \De\yb(rem).
\eeq
$\De\yb(dom)$ represents an asymptotic expansion including constant
terms for a high top-quark mass, and $\De\yb(rem)$ summarizes the remainder,
which vanishes by definition for $\Mt\to\infty$. For $\De\yb(dom)$ we
obtain
\beqar
\De\yb(dom )& = & \frac{\alpz}{12\pi c_0^2}\left[
\frac{3}{s_0^2}t + \frac{17-16s_0^2}{2s_0^2}\log(t)
+\frac{16}{9}(27-72s_0^2+52s_0^4)C_1
+\frac{6c_0^2}{s_0^2}(2-s_0^2)^2C_2
\right.\nn\\
&&\hphantom{\frac{\alpz}{8\pi c_0^2}}\left.
+\frac{12c_0^6}{s_0^2}(3-s_0^2)C_3
+\left(1-\frac{1}{2s_0^2}\right)(33-44s_0^2+12s_0^4)\logx(c_0^2)
\right.\nn\\
&&\hphantom{\frac{\alpz}{8\pi c_0^2}}\left.
+\left(\frac{13}{2s_0^2}-13+6s_0^2\right)\log(c_0^2)
+\frac{25}{3s_0^2}+\frac{392}{9}-119s_0^2+\frac{680s_0^4}{9}
\right],
\label{dybdom}
\eeqar
where we have used the shorthand
\beq
t = \frac{\Mt^2}{\MZ^2}.
\eeq
Since the full analytic form of the remainder is not very illuminating,
$\De\yb(rem)$ is given in \refapp{dybrem}. Here, we just give the
asymptotic expansion of $\De\yb(rem)$ up to $\O(t^{-2})$,
\beqar
\De\yb(rem )& = & \frac{\alpz}{48\pi s_0^2 c_0^2}\left[
(21-20s_0^2)(5-6s_0^2)\log\left(\frac{t}{c_0^2}\right)
+(3-4s_0^2)(33-44s_0^2+12s_0^4)\logx(c_0^2)
\right.\nn\\
&&\hphantom{\frac{\alpz}{48\pi s_0^2 c_0^2}}\left.
+\frac{2601}{20}-\frac{33259s_0^2}{90}+\frac{1004s_0^4}{3}-96s_0^6
\right] \; \frac{1}{t} \nn\\[.3em]
&& + \frac{\alpz}{240\pi s_0^2 c_0^2}\left[
(637-2244s_0^2+2630s_0^4-1020s_0^6)\log\left(\frac{t}{c_0^2}\right)
\right.\nn\\[.2em]
&&\hphantom{+\frac{\alpz}{240\pi s_0^2 c_0^2}}\left.
+(3-4s_0^2)(1-2s_0^2)(101-128s_0^2+30s_0^4)\logx(c_0^2)
\right.\nn\\[.2em]
&&\hphantom{+\frac{\alpz}{240\pi s_0^2 c_0^2}}\left.
+\frac{42123}{70}-\frac{303763s_0^2}{126}+\frac{31574s_0^4}{9}
-2178s_0^6+480s_0^8
\right] \; \frac{1}{t^2} \nn\\[.3em]
&& +\;\O\left(\frac{\log(t)}{t^3}\right).
\label{deyb}
\eeqar
Combining the results of (\ref{dybdom}) and (\ref{deyb}), the asymptotic
expansion of $\De\yb$ reads numerically
\beq
\begin{array}[b]{l}
\De\yb = \biggl( 3.47\,t \;+\; 7.70\,\log(t) \;-\; 17.69 \\
\phantom{\De\yb = \biggl(}
+\; 17.13\,\log(t)/t \;-\; 2.40/t
\;+\; 14.26\,\log(t)/t^2 \;+\; 7.45/t^2 \\
\phantom{\De\yb = \biggl(}
+\; \O\left(\log(t)/t^3\right)
\biggr) \;\times\; 10^{-3}.
\earr
\eeq
In \refta{dybtab}, we compare the exact values of $\De\yb$ with
the leading approximation $\De\yb^{\rm lead}$ and its
asymptotic expansions $\De\yb^{(k)}$ where terms of $\O(\log(t)/t^k)$
are neglected. In particular, $\De\yb^{\rm lead}$ contains only the $t$-
and $\log(t)$-terms of (\ref{dybdom}), $\De\yb^{(1)}$ is identical with
$\De\yb(dom)$, and the numerical values of $\De\yb^{(2)}$, $\De\yb^{(3)}$
are obtained from (\ref{deyb}).
We find that including only the
$t$- and $\log(t)$-terms of (\ref{dybdom}), i.e.\ $\De\yb^{\rm lead}$,
or adding in addition the constant term to obtain $\De\yb^{(1)}$,
is not a sufficient approximation for $\De\yb$, as can be clearly seen
in \refta{dybtab}. Therefore, we explicitly give further subleading
terms in (\ref{deyb}).
\btab \bce
$\begin{array}{|c|c|c|c|c|c|}
\hline
\Mt/\GeV & \De\yb^\onel/10^{-3} &
\De\yb^{\rm lead}/10^{-3} & \De\yb^{(1)}/10^{-3} &
\De\yb^{(2)}/10^{-3} & \De\yb^{(3)}/10^{-3}
\\ \hline
120 &  4.08 & 10.25 & -7.44 & -3.40 &  1.70 \\ \hline
160 & 10.14 & 19.36 &  1.67 &  7.15 &  9.62 \\ \hline
180 & 13.73 & 24.01 &  6.32 & 11.69 & 13.46 \\ \hline
240 & 26.51 & 38.97 & 21.28 & 25.73 & 26.46 \\ \hline
\earr$
\caption{Comparison of $\De\yb$ with the leading approximation,
$\De\yb^{\rm lead}$, and its asymptotic expansions
$\De\yb^{(k)}$ where terms of $\O(\log(t)/t^k)$ are neglected.}
\label{dybtab}
\ece
\etab

For a vanishing top-quark mass, $\Mt$, the parameter $\De\yb$ of course
coincides with $\De y_\Pd$ as given in (\ref{dyud}).

\section{Other important corrections}
\label{otherrcs}

\subsection{Leading two-loop top-corrections}
\label{top2l}

\btab \bce
\arraycolsep 6pt
$$\begin{array}{|c||c|c|c||c||c|c|c|}
\hline
\Mt/\GeV
& \multicolumn{3}{c||}{\De x^{\rm 1-loop}/10^{-3}}
& \De x\vert^\QCD_{\rm top,2l}/10^{-3}
& \multicolumn{3}{c|}{\De x\vert^{\rm weak}_{\rm top,2l}/10^{-3}} \\
\cline{2-4} \cline{6-8}
& \multicolumn{3}{c||}{\MH/\GeV=100,300,1000}
&& \multicolumn{3}{c|}{\MH/\GeV=100,300,1000} \\
\hline\hline
120 & ~~7.49~ & ~~6.71~ & ~~5.44~ & -0.48 & ~-0.02 & ~-0.05 & ~-0.06 \\ \hline
160 & ~11.98~ & ~11.20~ & ~~9.93~ & -0.86 & ~-0.05 & ~-0.13 & ~-0.19 \\ \hline
180 & ~14.48~ & ~13.70~ & ~12.43~ & -1.09 & ~-0.07 & ~-0.20 & ~-0.30 \\ \hline
240 & ~23.22~ & ~22.45~ & ~21.17~ & -1.94 & ~-0.13 & ~-0.52 & ~-0.92 \\ \hline
\end{array}$$
$$\begin{array}{|c||c||c||c|c|c|}
\hline
\Mt/\GeV & \De\yb^{\rm 1-loop}/10^{-3} & \De\yb\vert^\QCD_{\rm top,2l}/10^{-3}
& \multicolumn{3}{c|}{\De\yb\vert^{\rm weak}_{\rm top,2l}/10^{-3}} \\
\cline{4-6}
&&& \multicolumn{3}{c|}{\MH/\GeV=100,300,1000} \\
\hline\hline
120 & ~4.08 & -0.52 & ~~0.04~~ & ~~0.04~~ & ~~0.09~~ \\ \hline
160 & 10.14 & -0.92 & ~~0.15~~ & ~~0.12~~ & ~~0.22~~ \\ \hline
180 & 13.73 & -1.16 & ~~0.24~~ & ~~0.19~~ & ~~0.33~~ \\ \hline
240 & 26.51 & -2.07 & ~~0.83~~ & ~~0.62~~ & ~~0.86~~ \\ \hline
\end{array}$$
\caption{Leading two-loop top-corrections to $\De x$ and $\De\yb$ in
comparison with one-loop results.
``QCD'' corresponds to the terms of order $\O(\alps\al t)$,
``weak'' to the ones of order $\O(\al^2t^2)$.}
\label{t2ltab}
\ece \etab
The explicit expressions for the parameters $\De x$, $\De y$, $\eps$,
$\De\yb$, which have been given in \citere{di94} and \refse{anresm},
are valid in one-loop approximation, i.e.\ in $\O(\al)$. Although
up to now no complete two-loop calculation for the \PZO\ and $\mu^\pm$
decay widths exists, the leading two-loop top-corrections have already
been presented in the literature. QCD corrections of order $\O(\alps\al t)$
were given in \citere{dj87} and \citere{fl92} for the $\rho$-parameter
and $\Gb$, respectively. Moreover, corrections of order
$\O(\al^2 t^2)$ to the $\rho$-parameter for arbitrary Higgs mass were
presented in
\citeres{ba92,fl93} and to $\Gb$ in \citere{fl93}. It turns out that
these leading heavy-top contributions can be naturally embedded into
the parameters $\De x$, $\De y$, $\eps$, $\De\yb$. The $\rho$-parameter
enters merely $\De x$, and the process-specific heavy-top corrections
to $\Gb$ yield only contributions to $\De\yb$, while $\De y$ and $\eps$
are not affected. Following \citere{fl93}, we define
\beq
x_t \equiv \frac{\sqrt{2}G_\mu \Mt^2}{16\pi^2}
    = \frac{\alpz}{16\pi s_0^2c_0^2}\,t.
\eeq
For the higher-order top-effects on $\De x$ and $\De\yb$ we finally
obtain
\beqar
\De x\Big\vert_{\rm top,2l} &=&
9x_t^2 \,+\, 3x_t^2\rho^{(2)}(\Mt/\MH) \,+\, 3x_t\de^\QCD, \\
\De\yb\Big\vert_{\rm top,2l} &=&
12x_t^2 \,+\, 4x_t^2\tau_\Pb^{(2)}(\Mt/\MH) \,+\, 4x_t\de_\Pb^\QCD,
\eeqar
where the first $x_t^2$ terms on the r.h.s.\ represent reducible
(``squared one-loop'') contributions. The functions $\rho^{(2)}$
and $\tau_\Pb^{(2)}$, which depend on the ratio $\Mt/\MH$, can be
explicitly found in \citere{fl93}.
The QCD parts simply read from \citere{dj87}
\beq
\de^\QCD = -\frac{2\pi^2+6}{9} \frac{\alps}{\pi} =
-2.860\frac{\alps}{\pi}
\eeq
and \citere{fl92}
\beq
\de_\Pb^\QCD = -\frac{\pi^2-3}{3} \frac{\alps}{\pi} =
-2.290\frac{\alps}{\pi}.
\eeq
Both QCD contributions lead to a {\it screening} of the $\O(\al t)$
correction at one loop. In order to illustrate the influence of
these leading two-loop effects, we compare $\De x\vert_{\rm top,2l}$
and $\De\yb\vert_{\rm top,2l}$ in \refta{t2ltab} with the one-loop
results for $\De x$ and $\De\yb$ for various Higgs and top-quark
masses, respectively.
We note that the experimental accuracy in $\De x$ and $\De\yb$ is of the
order $5\times 10^{-3}$ (compare \refse{anexda}), which has to be
compared with the order $1\times 10^{-3}$ and $0.3\times 10^{-3}$ for the
$\O(\alps\al t)$ and $\O(\al^2 t^2)$ terms (for $\Mt\approx 175\GeV$),
respectively. Consequently, the (weak) $\O(\al^2 t^2)$ correction
turns out to be negligible.

\subsection{Finite-mass corrections}

In the results of the previous sections all fermions except for
the top-quark have been assumed to be massless. While this
approximation is obviously excellent for the leptons and the quarks of
the first and second fermion generation,
the finite-mass effects of the b-quark can reach the order
of the loop corrections even at the \PZO-mass scale. Consequently,
we include the $\O(\Mb^2/\MZ^2)$ correction to \Zbb,
which is simply given by
\beq
\de\Gb\Big\vert_{\mass} =
-\frac{\alpz}{8s_0^2c_0^2}\,N_{\rm C,b}\MZ\,\frac{\Mb^2}{\MZ^2}.
\label{finmas}
\eeq
Terms of order $\O(\Mb^4/\MZ^4)$ are completely negligible with
respect to the experimental and theoretical uncertainties.

\subsection{Higher-order QCD corrections}

The QCD corrections to the hadronic decays of the \PZO\ boson,
\Zqq, have been frequently discussed in the literature
(see e.g.\ \citeres{go91,dj87,fl92,kn90,ch90,kn89}).
Owing to finite-mass effects they are different for the vector and
axial-vector parts $\Ga_{\rm V,q}$ and $\Ga_{\rm A,q}$, respectively.
They are given by
\beqar
\de\Gq\Big\vert_{\QCD} &=& \phantom{+}
\Ga_{\rm V,q}\left\{
12\frac{\alps}{\pi}\frac{m_\Pq^2}{\MZ^2}\right\} \nn\\
&&+\Ga_{\rm A,q}\left\{
-6\frac{\alps}{\pi}\frac{m_\Pq^2}{\MZ^2}
\left[1+2\log\left(\frac{m_\Pq^2}{\MZ^2}\right)\right]
\pm \frac{1}{3}\left(\frac{\alps}{\pi}\right)^2
{\rm I}\left(\frac{1}{4t}\right)\right\},
\label{qcd}
\eeqar
where the ``$\pm$'' refers to u/d-type quarks, respectively, and
\beq
\Ga_{\rm V,q} = \frac{\alpz}{16s_0^2c_0^2}\,\MZ\,
(1-4s_0^2\vert Q_\Pq\vert)^2, \qquad
\Ga_{\rm A,q} = \frac{\alpz}{16s_0^2c_0^2}\,\MZ.
\eeq
The $m_\Pq$-dependent corrections in (\ref{qcd}) are
only relevant for q=b and have been taken from \citere{kn90,ch90}.
The full analytical expression for the function ${\rm I}(x)$ was
presented in \citere{kn89}. Instead, we use the excellent approximation
\beq
{\rm I}\left(\frac{1}{4t}\right) =
\frac{37}{4}+3\log(t)-0.26\,t^{-1}-0.04t^{-2}+\O(t^{-3}),
\eeq
which can also be found there.

\section{Analysis of the experimental data.}
\label{anexda}

\subsection{Input data.}
\label{data}

For our analysis we use the experimental data presented at the
Glasgow Conference \cite{sc94},
\beqar
\MZ &=& 91.1888 \pm 0.0044 \GeV, \nn\\
\GT &=& 2497.4 \pm 3.8 \MeV, \nn\\
R &=& \Gh/\Gl = 20.795 \pm 0.040, \nn\\ \disp
\si_{\rm h} &=& \frac{12\pi\Gl\Gh}{\MZ^2\Ga^2_{\rm T}} = 41.49\pm 0.12 \nba.
\label{scdata}
\eeqar
We take into account the correlation matrix for $\GT$,
$R$, and $\si_{\rm h}$,
\beq
\begin{array}{|c||c|c|c|c|}
\hline      & \GT    & R    & \si_{\rm h} \\ \hline\hline
\GT         & ~~1.00 & 0.01 & -0.11       \\ \hline
R           & ~~0.01 & 1.00 & ~~0.13      \\ \hline
{}~~~\si_{\rm h}~~~ & ~~-0.11~~ & ~~0.13~~ & ~~~~1.00~~       \\ \hline
\earr
\label{corr}
\eeq
since the other correlations (to $\MZ$ and $A^\Pl_{\rm FB}$)
are negligible. From (\ref{scdata}) and (\ref{corr}) one derives
\beqar
\Gl &=& 83.96 \pm 0.18 \MeV,\nn\\
\Gh &=& 1746  \pm 4 \MeV.
\eeqar
{}From the measured value of
\beq
R_{\rm bh} = \frac{\Gb}{\Gh} = 0.2202 \pm 0.0020,
\eeq
one then obtains
\beq
\Gb = 384.5 \pm 3.6 \MeV.
\eeq
{}From all asymmetries $(A^\Pl_{\rm FB}, A^\tau_{\rm pol}, A^\Pe,
A^\Pb_{\rm FB}, A^\Pc_{\rm FB})$
measured at LEP one deduces
\beq
\swbar^2 (\LEP) = 0.2321 \pm 0.0004.
\label{swlep}
\eeq
Upon including the SLD result on $A_{\rm LR}(\SLD)$ one has
\beq
\swbar^2 (\LEP+\SLD) = 0.2317 \pm 0.0004.
\eeq
In \citere{di94}, we have shown the numerical results using both
values for $\swbar^2$. Here, we restrict ourselves to the LEP result
(\ref{swlep}) for numerical evaluations. The results to be obtained upon
including the SLD value for $\swbar^2$ can be essentially inferred from
our previous analysis. Finally, we have
\beq
\frac{\MWpm}{\MZ} ({\rm UA2+CDF}) = 0.8798\pm0.0020.
\eeq

Concerning the input value of $\alpz$, there is an experimental
uncertainty due to the data on $\Pep\Pem\to\Pq\bar\Pq\to$ hadrons,
in particular in the low-energy region which strongly affects the
value of the dispersion integral employed when calculating $\alpz$.
The value \cite{alpz1} of
\beq
\alpz^{-1} = 128.87 \pm 0.12
\label{alpz1}
\eeq
has been supplemented by two new evaluations recently,
yielding \cite{sw94}
\beq
\alpz^{-1} = 129.08 \pm 0.10;
\eeq
and \cite{ma94}
\beq
\alpz^{-1} = 129.01 \pm 0.06.
\eeq
We also note that the above values are consistent with the
results
\beq
\alpz^{-1}= \left\{
\begin{array}{llll}
128.90 & \pm & 0.06, & \quad\mbox{\citere{ne94},} \\
128.96 & \pm & 0.03, & \quad\mbox{\citere{ge94},}
\end{array} \right.
\eeq
based on the experimental data on $\Pep\Pem \to$ hadrons in the
resonance regions in addition to stronger theoretical assumptions.
In our analysis we will use the value (\ref{alpz1}) of \citere{alpz1}
and indicate what will happen if this value is changed by the given
uncertainty.

In addition to $\alpz$, we will use
\beq
\GF=1.16639 (2) \cdot 10^{-5} \GeV^{-2},
\eeq
$\MZ$ from (\ref{scdata}), the strong coupling constant \cite{alps}
\beq
\alps = 0.118\pm0.007,
\label{alps}
\eeq
and the corresponding ``on-shell'' mass \cite{ch90} of the b-quark,
\beq
\Mb = 4.5\GeV,
\eeq
as input parameters.

\subsection{Determination of the parameters from experiment}
\label{exres}

In \refses{lepton} and \ref{arbfer}, we presented the explicit formulae
for the observables $\swbar^2$, $\MWpm/\MZ$, and $\Ga_\Pf$ in terms of the
effective parameters. In view of the experimental uncertainties
of these observables it is completely sufficient to consider the
contributions of the parameters $\De x$, $\De y$, $\eps$, $\De y_\Pf$,
$\de$, $\de_\Pq$ to the observables in linear order only, rendering our
investigation very transparent. Hence, we write
\beqar
\swbar^2 &=&
s_0^2\left[1+\si_x\De x+\si_y\De y+\si_\eps\eps+\si_\de\de\right],
\nn\\[.3em] \disp
\frac{\MWpm}{\MZ} &=&
c_0\left[1+\mu_x\De x+\mu_y\De y+\mu_\eps\eps+\mu_\de\de\right],
\nn\\[.3em]
\Gl &=&
\Gl^{(0)}\left[1+\ga^\Pl_x\De x+\ga^\Pl_y\De y+\ga^\Pl_\eps\eps
+\ga^\Pl_\de\de\right],
\nn\\[.3em]
\Ga_\nu &=&
\Ga_\nu^{(0)}\left[1+\ga^\nu_x\De x+\ga^\nu_y\De y
+\ga^\nu_{y_\nu}\De y_\nu\right],
\nn\\[.3em]
\Gq &=&
\Gq^{(0)}\left[1+\ga^\Pq_x\De x+\ga^\Pq_y\De y+\ga^\Pq_\eps\eps
+\ga^\Pq_{y_\Pq}\De y_\Pq+\ga^\Pq_\de\de+\ga^\Pq_{\de_\Pq}\de_\Pq\right]
+\de\Gq\Big\vert_\mass
+\de\Gq\Big\vert_\QCD,
\nn\\
\label{obsalllin}
\eeqar
where the coefficients $\si_i$, $\mu_i$, $\ga^\Pf_i$ can be easily read
from (\ref{obsleplin}), (\ref{obsnulin}), (\ref{obshadlin}), and the
lowest-order contributions $s_0^2$, $c_0$, $\Ga_\Pf^{(0)}$ are defined in
(\ref{s02}), (\ref{obsleplow}), (\ref{obsnuhadlow}). Here,
$\de\Gq\vert_\mass$ and $\de\Gq\vert_\QCD$ denote the finite-mass
and higher-order QCD corrections given in (\ref{finmas}) and
(\ref{qcd}), respectively. Based on (\ref{obsalllin}), the hadronic
width, $\Gh$, and the total width, $\GT$, take the form
\beqar
\Gh &=& 2\Gu+2\Gd+\Gb \nn\\[.3em]
&=&
\Gh^{(0)}\left[1+\ga^{\rm h}_x\De x+\ga^{\rm h}_y\De y+\ga^{\rm h}_\eps\eps
+\ga^{\rm h}_{y_\Pb}\De\yb+\ga^{\rm h}_{\yh}\De\yh
+\ga^{\rm h}_\de\de_{\rm h}\right]
+\de\Gb\Big\vert_\mass
+\de\Gh\Big\vert_\QCD,
\nn\\[.6em]
\GT &=& 3\Gl+3\Ga_\nu+2\Gu+2\Gd+\Gb \nn\\[.3em]
&=&
\GT^{(0)}\left[1+\ga^{\rm T}_x\De x+\ga^{\rm T}_y\De y+\ga^{\rm T}_\eps\eps
+\ga^{\rm T}_{y_\Pb}\De\yb+\ga^{\rm T}_{\yh}\De\yh
+\ga^{\rm T}_\de\de_{\rm T}\right]
+\de\Gb\Big\vert_\mass
+\de\Gh\Big\vert_\QCD.
\nn\\
\label{obsalllin2}
\eeqar
Since the partial decay widths \Zuudd\ cannot be measured separately,
the parameters $\De\yu$ and $\De\yd$ cannot be resolved by
experiment. They appear only in the combination
\beq
\De\yh = \frac{1}{2}(\De y_\Pu+\De y_\Pd)
+\frac{s_0^2}{6c_0^2}(\De y_\Pd-\De y_\Pu)
\eeq
in the hadronic and total widths in (\ref{obsalllin2}).
The parameters $\de_\Ph$ and $\de_{\rm T}$ summarize
the contributions of $\de$ and $\de_\Pq$.

The linearized equations (\ref{obsalllin}) and (\ref{obsalllin2}) may
now be used to extract the parameters $\De x$ etc.\ from the
experimental data on $\swbar^2$ etc.
Conversely, employing the formulae for the standard values of
the effective parameters $\De x$
etc.\ presented in the previous sections and \citere{di94}, the
linearized equations
(\ref{obsalllin}) and (\ref{obsalllin2}) can be used to evaluate the
complete standard model prediction for the observables at one loop.
In order to obtain also the leading two-loop top-corrections in $\De x$
and $\De\yb$ (see \refse{top2l}) correctly, (\ref{obsalllin}) and
(\ref{obsalllin2}) have to be completed by the corresponding quadratic
terms.

In our analysis of the experimental data, we employ a {\it two-step
procedure}. In a {\it first step}, we determine the experimental
values of the six parameters
\beq
\De x,\De y, \eps; \quad \De\yb, \De \yh, \De y_\nu
\eeq
from the six experimental data
\beq
\swbar^2, \MWpm/\MZ, \Gl; \quad \Gh, \GT, \Gb,
\eeq
where the contributions of $\de$, $\de_\Pq$, $\de\Gb\Big\vert_{\mass}$,
and $\de\Gh\Big\vert_{\QCD}$ are taken from theory.%
\footnote{Note that $\de\Gh\Big\vert_{\QCD}$ according to (\ref{qcd})
(smoothly) depends on the top mass, $\Mt$, in order $\O(\alps^2)$.
For the evaluations, we choose $\Mt=175\GeV$ and remark that e.g.\ a change
in $\Mt$ of $\pm 50\GeV$ leads to a variation of at most $0.6\times
10^{-3}$ in $\De\yb^{\rm exp}$, which is very much smaller than the
experimental error.}
The results for $\De x, \De y,\eps$ only depend on the input for
$\Gl, \swbar^2, \MWpm/\MZ$. They are shown in
\reffis{xefig}a)-\ref{eyfig}a).
A comparison of these 83 \% C.L. contours in the respective planes
with our previous ones \cite{di94} shows a decrease in the sizes of the
ellipses due to a somewhat decreased experimental error. The main
conclusion from these results has been presented in \citeres{bi93,di94}:
the data have reached such a high accuracy {\it that contributions beyond
the fermion-loop contributions to the $\ga$, \PWpm\ and \PZO\ propagators
are clearly required}. In particular, the data require a significant
contribution to $\De y$ which in the standard electroweak theory
is provided
by vertex corrections to the $\PWpm\Pf\Pf^\prime$ and $\PZO\Pf\Pf$
vertices which involve trilinear vector-boson self-couplings.
On the other hand, $\De y$ is practically independent \cite{di94a}
of the mass of the Higgs scalar and the concept of the Higgs mechanism.
The experimental data on $\eps$ and $\De x$ are well approximated by
\beq
\De x\approx\De x^\fer, \qquad \eps\approx\eps^\fer.
\eeq
The numerical values for the parameters $\De x^{\rm exp}$,
$\De y^{\rm exp}$, $\eps^{\rm exp}$, shown in
\reffis{xefig}a)-\ref{eyfig}a), are given by
\beq
\begin{array}[b]{rcrcrcrcrl}
\De x^{\rm exp}     &= (&  9.8 &\pm& 4.7 &\pm& 0.2 &\pm& 0 &)
\times 10^{-3},\\
\De y^{\rm exp}     &= (&  4.6 &\pm& 4.9 &\pm& 0.2 &\pm& 0 &)
\times 10^{-3},\\
\eps^{\rm exp}      &= (& -6.1 &\pm& 2.0 &\mp& 0.7 &\pm& 0 &)
\times 10^{-3},
\earr
\label{gb6xye}
\eeq
where the first error is statistical (1 $\si$), the second due to
the deviation by replacing $\alpz^{-1}\to\alpz^{-1}\pm\de\alpz^{-1}$
according to (\ref{alpz1}), and the third due to
$\alps\to\alps\pm\de\alps$ according to (\ref{alps}).
In \reffis{xefig}-\ref{eyfig}, the shift of the (center of the) ellipses
as a result of the changes $\alpz^{-1}\to\alpz^{-1}+\de\alpz^{-1}$ and
$\alps\to\alps+\de\alps$ is indicated by an arrow in the upper left-hand
corner.
Note that the uncertainties in $\De x^{\rm exp}$ etc.\ induced by
$\de\alpz^{-1}$ and $\de\alps$ are mainly due to the subtractions
of the ``lowest-order'' contributions, $s_0^2$, $c_0$, $\Gl^{(0)}$ etc.,
and the mass and QCD corrections, $\de\Gq\Big\vert_\mass$ and
$\de\Gq\Big\vert_\QCD$, in (\ref{obsalllin}) and (\ref{obsalllin2}),
respectively.
The theoretical predictions, $\De x^{\rm th}$ etc., are only
influenced via higher orders and lead to entirely negligible shifts in
the figures.

In \reffis{bxfig}a)-\ref{befig}a),
we show the results for $\De\yb$ in conjunction with $\De x,\De y,\eps$.
As expected from the known discrepancy between experiment and theory
in $\Gb$, the data clearly indicate a value of $\De\yb$
which does not show the expected enhancement due to a large mass
of the top-quark.
Numerically, $\De\yb^{\rm exp}$ is given by
\beq
\De\yb^{\rm exp} = ( -9.5 \pm 8.2 \pm 0.0 \pm 1.8 ) \times 10^{-3}.
\label{gb6b}
\eeq
In \reffis{xefig}-\ref{befig}, we also show the results of taking into
account fermion loops only, obviously corresponding to $\De\yb=0$ as
$\De\yb$ gets no fermion-loop contributions. We have also indicated for
comparison the value of $\De\yd$ in \reffis{bxfig}-\ref{befig} by an
arrow, which corresponds to $\De\yb$ for $\Mt\to 0$. It
seems that the data on $\Gb$ prefer a theoretical value for $\De\yb$
which does not contain the effect of the $\Mt$-dependent vertex correction.

For the remaining parameters, $\De\yh$ and $\De y_\nu$, which are not
shown in figures, we find
\beq
\begin{array}[b]{rcrcrcrcrl}
\De\yh^{\rm exp}    &= (&  1.2 &\pm& 2.8 &\pm& 0.0 &\pm& 1.9 &)
\times 10^{-3},\\
\De y_\nu^{\rm exp} &= (&  1.3 &\pm& 7.7 &\pm& 0.0 &\pm& 0.0 &)
\times 10^{-3}.
\earr
\label{yhnexp}
\eeq
These values are seen to be consistent with the theoretical predictions
\beqar
\De\yh^{\rm th}    &=& -1.37 \times 10^{-3}, \nn\\
\De y_\nu^{\rm th} &=& -3.05 \times 10^{-3}.
\label{yhnth}
\eeqar

We note that the uncertainty in $\alpz$ strongly influences the
parameter $\eps$ and accordingly the determination of the Higgs mass, while
it is fairly irrelevant for the remaining parameters. The uncertainty in
$\al_s$ is reflected in the determination of the (hadronic) parameters
$\De\yb$ and $\De\yh$.

We turn to the {\it second step} of our analysis. As noted in \refse{anresm},
the theoretical predictions for $\De\yh$ and $\De y_\nu$ are
actually on the same footing as the theoretical predictions for the
fermion-loop contributions to the $\ga, \PWpm$ and
\PZO\ propagators. Both predictions involve vector-boson fermion couplings
only, and are (consequently) independent of the (empirically unknown)
vector-boson self-couplings.
Motivated by the consistency between theory and experiment for $\De\yh$,
$\De y_\nu$ in (\ref{yhnth}) and (\ref{yhnexp}), we now impose the
assumption that the process-specific vertex corrections $\De\yh$, $\De y_\nu$
are given by the standard values (\ref{yhnth}).
Accordingly, the number of six free (fit) parameters, thus,
being reduced to four, $\De x,\De y,\eps$ and $\De\yb$.
Concerning the experimental input, we will discriminate between six
$(\Gl,\swbar,\MWpm/\MZ,\Gh,\GT,\Gb)$ and five
$(\Gl,\swbar,\MWpm/\MZ,\Gh,\GT)$ input data.
This discrimination allows us to analyze the influence of the data for
$\Gb$ on the results for $\De x$, $\De y$, $\eps$, and $\De\yb$.
We refer to these cases by ``with $\Gb$'' and ``without
$\Gb$'', respectively.

We find
\beq
\begin{array}[b]{rcrcrcrcrl}
\mbox{with}\;\; \Gb: \qquad
\De x^{\rm exp}     &= (&  9.6 &\pm& 4.7 &\pm& 0.2 &\mp& 0.0 &)
\times 10^{-3},\\
\De y^{\rm exp}     &= (&  5.6 &\pm& 4.8 &\pm& 0.2 &\pm& 0.4 &)
\times 10^{-3},\\
\eps^{\rm exp}      &= (& -5.2 &\pm& 1.8 &\mp& 0.7 &\pm& 0.3 &)
\times 10^{-3},\\
\De\yb^{\rm exp}    &= (& -3.3 &\pm& 5.9 &\pm& 0.0 &\pm& 5.8 &)
\times 10^{-3}.
\earr
\label{gb4xyeb}
\eeq
A comparison of these numerical results%
\footnote{Note added: While proofreading the final version of the
	present paper, we obtained the final
	draft of the preprint CERN-TH 7536/94 by G.\
	Altarelli, R.\ Barbieri, and F.\ Caravaglios,
	which contains an update of their analysis of the \PZO\ data in
	terms of the parameters $\eps_{1,2,3}$,$\eps_\Pb$ \cite{alta}.
	These parameters are related to ours by
	$$\barr{l}
	\eps_1=\De x-\De y+4s_0^2\de=\De x-\De y+0.2\times 10^{-3},\\
	\eps_2=-\De y+2s_0^2\de=-\De y+0.1\times 10^{-3},\\
	\eps_3=-\eps+\de=-\eps+0.2\times 10^{-3},\\
	\eps_\Pb=-\De\yb/2+4s_0^2[(2s_0^2-3)\de+s_0^2\de_\Pb]
		   /(9-6s_0^2)=-\De\yb/2-0.1\times 10^{-3}.
	\earr$$
	Repeating our analysis for precisely the same experimental data
	used there ($\swbar^2(\LEP+\SLD)=0.2317\pm0.0004$,
	$R_{\rm bh}=0.2192\pm0.0018$),
	we reproduced the values for $\eps_{1,2,3}$,
	within $0.1\times 10^{-3}$ and the one for $\eps_\Pb$
	within $1\times 10^{-3}$. In view of the differences between the
	two analyses (lowest-order and QCD subtractions etc.), a
	deviation in $\eps_\Pb$ of this order is to be expected.
	We note that the dominance of the fermion loops is clearly
	visible in our set of parameters, where
	$\De x\approx\De x^\fer$ in the standard electroweak theory,
	while this fact is concealed in the linear combination,
	$\eps_1$.}
(\ref{gb4xyeb}) with the ones of
the six-parameter analysis (\ref{gb6xye}), (\ref{gb6b}) does not reveal
dramatic differences in $\De x$, $\De y$, $\eps$. The experimental
result for $\De\yb$ is shifted into the direction of the theoretical
prediction by roughly one standard deviation.
This is also evident from comparing
\reffis{xefig}a)-\ref{befig}a) with \reffis{xefig}b)-\ref{befig}b).
The sizes of the 83\% C.L.\ ellipses, as expected, are slightly
decreased in \reffis{xefig}b)-\ref{befig}b) due to the smaller number of
four (fit) parameters. On the other hand, the dependence on $\alps$ is
somewhat stronger in the four-parameter fit.
As in \reffis{xefig}a)-\ref{befig}a) for the six-parameter analysis, in
\reffis{xefig}b)-\ref{befig}b) we indicate the $\alpz$- and
$\alps$-uncertainties of the experimental ellipses by arrows in the
upper left-hand corner. The long arrow for $\de\alps$ in
\reffis{bxfig}b)-\ref{befig}b) corresponds to the case
``without $\Gb$'', where $\Gb$ is excluded from the fit. Here, we find
\beq
\begin{array}[b]{rcrcrcrcrl}
\mbox{without}\;\; \Gb: \qquad
\De x^{\rm exp}     &= (&  9.6 &\pm& 4.7 &\pm& 0.2 &\pm& 0.0 &)
\times 10^{-3},\\
\De y^{\rm exp}     &= (&  5.0 &\pm& 4.8 &\pm& 0.2 &\pm& 0.0 &)
\times 10^{-3},\\
\eps^{\rm exp}      &= (& -5.7 &\pm& 1.8 &\mp& 0.7 &\pm& 0.0 &)
\times 10^{-3},\\
\De\yb^{\rm exp}    &= (&  1.0 &\pm& 7.2 &\pm& 0.0 &\pm& 8.8 &)
\times 10^{-3}.
\earr
\label{nogb4xyeb}
\eeq
As expected, the main change in (\ref{nogb4xyeb}) relative to
(\ref{gb4xyeb}) occurs in $\De\yb$, which is moved upwards
in the direction of better agreement with the theoretical expectation for a
heavy top-quark.
We note that even upon excluding the data for $\Gb$, the agreement of
the fit with theory is not perfect. If we impose the additional
constraint of $\Mt\approx 175\GeV$, the value of $\Mt$ indicated by the
direct searches \cite{mt}, the theoretical prediction is at the edge of
the experimentally allowed region, rather than in the center.
Inspecting the $\alps$-dependence of the results, we see that the
agreement between experiment and theory for $\Mt\approx 175\GeV$ will be
considerably  improved if $\alps$ approaches higher values such as
$\alps+\de\alps$. This conclusion is in accordance with $\alps$-fits
performed by other groups.

\section{Conclusions}
\label{concl}

Upon having included the hadronic \PZO\ decays in our analysis, we
reemphasize our previous conclusions. The parameters $\De x$, $\De y$,
$\eps$, and $\De\yb$, introduced as a parametrization of $SU(2)$-symmetry
breaking in an effective Lagrangian, are empirically found to have a
magnitude typical for radiative corrections. This in itself is a major
triumph of the $SU(2)$ symmetry properties embodied in the standard
electroweak theory. Moreover, discriminating
between the pure fermion-loop corrections
(only dependent on the couplings of the fermions to the vector bosons) and
the remaining ``bosonic'' corrections (dependent on the non-Abelian
vector-boson couplings and the Higgs scalar), we have found that the
experiments have become
sufficiently accurate to require corrections beyond pure fermionic
vacuum-polarization effects. More specifically, the data require a
non-vanishing value of the bosonic contribution to
$\De y$ which in the standard model is induced by
(combined) vertex corrections at the
$\PWp\Pl\bar\nu$ ($\PWm\nu\bar\Pl$) and $\PZO\Pl\bar\Pl$ vertices.
These corrections depend on the non-Abelian structure of the
vector-boson sector of the theory.

The leading two-loop top-corrections to the $\rho$-parameter and \Zbb\
are naturally absorbed by $\De x$ and $\De\yb$, respectively. For a top
mass of about $175\GeV$, the $\O(\alps\al t)$ terms are of the order
$1\times 10^{-3}$, the $\O(\al^2 t^2)$ terms of the order
$0.3\times 10^{-3}$, which have both to be confronted with the present
experimental error of the order of $5\times 10^{-3}$ in $\De x$ and
$\De\yb$.

Technically, in order to encourage experimentalists to carry out such an
analysis themselves with future precision data, we have given all
relevant analytic formulae explicitly
which are necessary to extract the parameters $\De x$, $\De y$ etc.\
from the data by a fitting procedure. Moreover, the standard-model
values for these parameters have been explicitly displayed in an
analytically compact form, which can also be easily evaluated numerically.

\section*{Acknowledgement}
One of the authors (D.S.) thanks Guido Altarelli for
useful discussions.

\appendix
\section*{Appendix}

\section{\boldmath{Remainder of $\De\yb$}}
\label{dybrem}

In \refse{anresm}, we have split the parameter $\De\yb$ into a
dominant and remainder part $\De\yb(dom)$ and $\De\yb(rem)$,
respectively, but we have only given the first few
asymptotic terms of the remainder. Recall that $\De\yb(rem)$ is
defined such that $\De\yb(dom)$ contains all contributions which
do not vanish for $t\to\infty$. Here, we present the full formula,
\beqar
\De\yb(rem )& = & \frac{\alpz}{24\pi s_0^2 c_0^2}\Biggl[
-\frac{23}{3}+\frac{71s_0^2}{9}+8s_0^4c_0^2
+\frac{3c_0^4(3-2s_0^2)}{t-c_0^2}
-\left(\frac{33}{2}-21s_0^2+4s_0^4\right)t
\nn\\
&&\hphantom{\frac{\alpz}{24\pi s_0^2 c_0^2}\Biggl[}\left.
-2(3-2s_0^2)t^2
-(17-16s_0^2)\log\left(\frac{t}{c_0^2}\right)
\right.\nn\\
&&\hphantom{\frac{\alpz}{24\pi s_0^2 c_0^2}\Biggl[}
+\frac{\log\left(t/c_0^2\right)}{(t-c_0^2)^2}
\left\{2c_0^6(3-4s_0^2)(5-2s_0^2)-4c_0^4(27-41s_0^2+12s_0^4)t
\right.\nn\\
&&\hphantom{\frac{\alpz}{24\pi s_0^2 c_0^2}\Biggl[+}\left.
        +2c_0^2(45-67s_0^2+18s_0^4)t^2-(15-10s_0^2-8s_0^4)t^3
        -6(1-2s_0^2)t^4\right\}
\nn\\
&&\hphantom{\frac{\alpz}{24\pi s_0^2 c_0^2}\Biggl[}\left.
+\left\{-30+82s_0^2-68s_0^4+16s_0^6+4(3-6s_0^2+2s_0^4)t
        -8s_0^2t^2\right\}\logx(t)
\right.\nn\\
&&\hphantom{\frac{\alpz}{24\pi s_0^2 c_0^2}\Biggl[}
+\left\{33-110s_0^2+100s_0^4-24s_0^6-3(7-12s_0^2+4s_0^4)t
\right.\nn\\
&&\hphantom{\frac{\alpz}{24\pi s_0^2 c_0^2}\Biggl[+\Biggl\{}\left.
        -6(1-2s_0^2)t^2\right\}\logx(c_0^2)
\nn\\
&&\hphantom{\frac{\alpz}{24\pi s_0^2 c_0^2}\Biggl[}
+2\left\{-2c_0^2(3-4s_0^2)(2-s_0^2)^2+4c_0^2(6-8s_0^2+3s_0^4)t
         -(9-10s_0^2)t^2
\right.\nn\\
&&\hphantom{\frac{\alpz}{24\pi s_0^2 c_0^2}\Biggl[+2\Biggl\{}\left.
         +4s_0^2t^3\right\} C_4(t)
\nn\\
&&\hphantom{\frac{\alpz}{24\pi s_0^2 c_0^2}\Biggl[}
+6\left\{-4c_0^6(3-s_0^2)+c_0^2(7-3s_0^2)(1-2s_0^2)t
         -(3-4s_0^2)t^2
\right.\nn\\
&&\hphantom{\frac{\alpz}{24\pi s_0^2 c_0^2}\Biggl[+2\Biggl\{}\left.
         -(1-2s_0^2)t^3\right\} C_5(t)
\Biggr].
\eeqar
The auxiliary functions $\logx(x)$, $C_4(t)$, $C_5(t)$ are explicitly
given in \refapp{aux}.

\section{Auxiliary functions}
\label{aux}

Here, we list the explicit expressions for the auxiliary functions
which have been used in \refse{anresm} and \refapp{dybrem}.
$\logx(x)$ is defined by
\beq
\logx(x) = \Re\left[\beta_x\log\left(\frac{\beta_x-1}
              {\beta_x+1}\right)\right], \quad\mbox{with}\;\;
             \beta_x=\sqrt{1-4x+i\epsilon}.
\eeq
The constants $C_1, C_2, C_3$ and the functions $C_4(t), C_5(t)$
are shorthands for the scalar three-point
integrals occuring in the process dependent parts of the decay \Zff,
\beqar
C_1 &=& \MZ^2\Re\left[C_0(0,0,\MZ^2,0,\MZ,0)\right] =
-\frac{\pi^2}{12} = -0.8225, \nn\\[.3em]
C_2 &=& \MZ^2\Re\left[C_0(0,0,\MZ^2,0,\MW,0)\right] =
\frac{\pi^2}{6}-\Re\left[\Li\left(1+\frac{1}{c_0^2}\right)\right]
= -0.8037, \nn\\[.3em]
C_3 &=& \MZ^2\Re\left[C_0(0,0,\MZ^2,\MW,0,\MW)\right]
= \Re\left[\log^2\left(\frac{i\sqrt{4c_0^2-1}-1}
                          {i\sqrt{4c_0^2-1}+1}\right)\right]
= -1.473, \nn\\
C_4(t) &=& \MZ^2\Re\left[C_0(0,0,\MZ^2,\Mt,\MW,\Mt)\right], \nn\\
C_5(t) &=& \MZ^2\Re\left[C_0(0,0,\MZ^2,\MW,\Mt,\MW)\right].
\label{cfctns}
\eeqar
The first three arguments of the $C_0$-function label the external
momenta
squared, the last three the inner masses of the corresponding vertex
diagram. All $C_0$-functions occuring in (\ref{cfctns}) follow from
the more general result
\beqar
C_0(0,0,s,m_1,m_0,m_1) & = & -\frac{1}{s} \Biggl[
 \Li\left(1-\frac{m_0^2}{m_1^2}\right)
-\Li\left(1+\frac{m_0^2}{m_1^2}x_2\right)
-\Li(1+x_2)
\nn\\[.3em]
&& \hspace{-5em}
+ \frac{1}{2}\log^2(-x_1)
+\sum_{\sigma=\pm 1} \left\{
\Li\left(1-x_1^\sigma x_2\right)
+\eta(-x_2,-x_1^\sigma) \log\left(1-x_1^\sigma x_2\right) \right\}
\Biggr],\hspace{2em}
\eeqar
with
\beq
x_1 = \frac{1+\beta}{1-\beta}, \quad
x_2 = \frac{s+\ie-m_1^2+m_0^2}{m_1^2-m_0^2}, \quad
\beta = \sqrt{1-\frac{4m_1^2}{s+\ie}}.
\eeq
The dilogarithm $\Li(x)$ and the $\eta$-function $\eta(x,y)$
are defined as usual,
\beqar
\Li(x) = -\int_0^x\,\frac{dt}{t}\,\log(1-t),
&&\qquad -\pi<{\mathrm arc}(1-x)<\pi,\\[.2em]
\eta(x,y) = \log(xy)-\log(x)-\log(y),
&&\qquad -\pi<{\mathrm arc}(x),\mathrm{arc}(y)<\pi.
\eeqar

\clearpage

\begin{figure}
\begin{center}
\begin{picture}(16,8.1)
\put(-3.0,-16.1){\includegraphics{xefig.ps}}
\put( 7.3,1.3){\footnotesize Fig.\ a)}
\put(14.7,1.3){\footnotesize Fig.\ b)}
\end{picture}
\end{center}
\caption{\protect{
83\% C.L.\ ellipse in the $\De x$-$\eps$-plane obtained in
a) a six-parameter analysis ($\De x,\De y,\eps,\De\yb,\De\yh,\De y_\nu$),
b) a four-parameter analysis ($\De x,\De y,\eps,\De\yb$)
of the data.
The full standard model predictions are shown for Higgs masses of
$100\GeV$ (dotted with diamonds), $300\GeV$ (long-dashed--dotted), $1\TeV$
(short-dashed--dotted) parametrized by the top-quark mass ranging from
$100$-$240\GeV$ in steps of $20\GeV$. The pure fermion-loop prediction is
also shown (short-dashed curve with squares) for the same top-quark masses.}}
\label{xefig}
\efi
\begin{figure}
\begin{center}
\begin{picture}(16,8.1)
\put(-3.0,-16.1){\includegraphics{xyfig.ps}}
\put( 7.3,1.3){\footnotesize Fig.\ a)}
\put(14.7,1.3){\footnotesize Fig.\ b)}
\end{picture}
\end{center}
\caption{83\% C.L.\ ellipse in the $\De x$-$\De y$-plane.
See also caption of \reffi{xefig}.}
\label{xyfig}
\efi
\begin{figure}
\begin{center}
\begin{picture}(16,8.1)
\put(-3.0,-16.1){\includegraphics{eyfig.ps}}
\put( 7.3,1.3){\footnotesize Fig.\ a)}
\put(14.7,1.3){\footnotesize Fig.\ b)}
\end{picture}
\end{center}
\caption{83\% C.L.\ ellipse in the $\eps$-$\De y$-plane.
See also caption of \reffi{xefig}.}
\label{eyfig}
\efi
\begin{figure}
\begin{center}
\begin{picture}(16,8.1)
\put(-3.0,-16.1){\includegraphics{bxfig.ps}}
\put( 7.3,1.3){\footnotesize Fig.\ a)}
\put(14.7,1.3){\footnotesize Fig.\ b)}
\end{picture}
\end{center}
\caption{83\% C.L.\ ellipse in the $\De\yb$-$\De x$-plane.
See also caption of \reffi{xefig}.
The short/long arrows for $\de\al_s$ in Fig.\ b)
correspond to the cases with/without $\Gb$, respectively.}
\label{bxfig}
\efi
\begin{figure}
\begin{center}
\begin{picture}(16,8.1)
\put(-3.0,-16.1){\includegraphics{byfig.ps}}
\put( 7.3,1.3){\footnotesize Fig.\ a)}
\put(14.7,1.3){\footnotesize Fig.\ b)}
\end{picture}
\end{center}
\caption{83\% C.L.\ ellipse in the $\De\yb$-$\De y$-plane.
See also caption of \reffi{xefig}.
The short/long arrows for $\de\al_s$ in Fig.\ b)
correspond to the cases with/without $\Gb$, respectively.}
\label{byfig}
\efi
\begin{figure}
\begin{center}
\begin{picture}(16,8.1)
\put(-3.0,-16.1){\includegraphics{befig.ps}}
\put( 7.3,1.3){\footnotesize Fig.\ a)}
\put(14.7,1.3){\footnotesize Fig.\ b)}
\end{picture}
\end{center}
\caption{83\% C.L.\ ellipse in the $\De\yb$-$\eps$-plane.
See also caption of \reffi{xefig}.
The short/long arrows for $\de\al_s$ in Fig.\ b)
correspond to the cases with/without $\Gb$, respectively.}
\label{befig}
\efi

\end{document}